\documentclass[conference]{IEEEtran}
\IEEEoverridecommandlockouts

\usepackage[T1]{fontenc}
\usepackage{cite}
\usepackage{amsmath,amssymb,amsfonts,amsthm}
\usepackage{graphicx}
\usepackage{textcomp}
\usepackage[dvipsnames]{xcolor}
\usepackage{braket}
\usepackage{subcaption}
\usepackage{float}
\usepackage{tikz}
\usepackage{quantikz}
\usepackage{pgfplots}
\usepackage{enumitem}
\usepackage{booktabs}
\usepackage{amssymb}
\usepackage[linesnumbered,ruled,noend]{algorithm2e}

\usepackage{hyperref}
\hypersetup{
    colorlinks = true,
    linkcolor = BlueViolet,
    citecolor = BlueViolet,
    filecolor = BlueViolet,
    urlcolor = BlueViolet,
}

\makeatother

\def\BibTeX{{\rm B\kern-.05em{\sc i\kern-.025em b}\kern-.08em
    T\kern-.1667em\lower.7ex\hbox{E}\kern-.125emX}}

\newtheorem{definition}{Definition}

\newtheorem{prop}{Proposition}

\newtheorem*{quantumping}{Quantum Ping (QPing)}
\newtheorem*{passivequantumping}{Passive Quantum Ping}
\theoremstyle{remark}
\newtheorem*{remark}{Remark}
\newcommand\blfootnote[1]{%
  \begingroup
    \renewcommand\thefootnote{}%
    \footnote{#1}%
    \addtocounter{footnote}{-1}%
  \endgroup
}

\begin{document}
\title{QPing: a Quantum Ping Primitive for Quantum Networks}

\author{\IEEEauthorblockN{
    Jorge Miguel-Ramiro*, Jessica Illiano$^{\dagger}$, Francesco Mazza$^{\dagger}$, Alexander Pirker*, 
    Julia Freund*,\\ 
    Angela Sara Cacciapuoti$^{\dagger}$, Marcello Caleffi$^{\dagger}$, and Wolfgang D\"ur*
    }

\IEEEauthorblockN{$^{*}$Universit\"at Innsbruck, Institut f\"ur Theoretische Physik, Technikerstra{\ss}e 21a, Innsbruck 6020, Austria\\ $^{\dagger}$University of Naples Federico II, Naples 80125, Italy}
}

\maketitle
\thispagestyle{plain}
\pagestyle{plain}

\begin{abstract}
We introduce the concept of Quantum Ping (QPing) as a diagnostic primitive for future quantum networks, designed to assess whether two or more end nodes can establish practical quantum entanglement with efficient resource consumption, limited overhead, and time-adaptive fidelity thresholds. Unlike classical ping, which probes network-layer connectivity through ICMP messages, our proposed quantum version is adapted to the unique features of quantum networks, where connectivity depends on the availability and quality of shared entanglement. We develop a formal framework for QPing and leverage different tools such as sequential hypothesis testing to probe quantum connectivity. We present several strategies, including active strategies, with path-based and segment-based variants, and passive strategies that utilize pre-shared entangled resources. QPing can serve as a flexible diagnostic building block for quantum networks, designed to work alongside fundamental network operations, while remaining suitable to different architectural and protocol design approaches.

\end{abstract}

\medskip

\begin{IEEEkeywords}
Entanglement, Quantum Networks, Quantum Communication, Quantum Routing, Quantum State Verification, Fidelity Witnessing
\end{IEEEkeywords}

\blfootnote{This work has been
submitted to the IEEE for possible publication. Copyright may be transferred without notice, after
which this version may no longer be accessible.}

\section{Introduction}
Quantum networks promise to introduce a new paradigm in communication, enabling unique and powerful applications such as distributed sensing \cite{Giovannetti_2011, Kessler2014, Sekatski2020, GiaWinCon-25}, quantum cryptography \cite{Gisin2002, Pirandola_2020}, and distributed quantum computation \cite{CiracDistributed, Hayashi15, Cacciapuoti2020}, among others \cite{jensen2025quantum}. These capabilities are expected to come together in a comprehensive, large-scale quantum network, linking users and cities, and ultimately giving rise to the quantum Internet \cite{Kimble2008, Wehner2018, CacCalVan-20, Gyongyosi_2022, Koji2023, rfc9583}.

Significant efforts have been dedicated to analyzing both the fundamental and advanced components required for a quantum network \cite{Pirker_2018, Kozlowski2019, Kozlowski_2020}. These studies encompass hierarchical stack models \cite{Pirker_2019, Dahlberg_2019, Illiano_2022} that structure network elements into distinct layers, facilitating modular development and clearer functionality. Key devices within these networks include quantum repeaters  \cite{Briegel_Repeaters, D_r_2016, Meter_2013, MR2023}, which extend communication ranges by overcoming quantum signal degradation, but also elements as routers \cite{Van_Meter_2013, Lee_2022,Shi_2024,Abane_2025} designed to manage efficient state connectivity across the network. Protocols such as those for entanglement distribution and manipulation \cite{Meignant2019, Navascues2020, LiXueLi-23, Miguel_Ramiro_2023, Fan2024} are critical for establishing and maintaining quantum entanglement between network nodes. Many of these elements exhibit parallels to the classical Internet functionalities, but are specifically tailored to address the unique challenges of quantum communication.

In this work, we consider a fundamental component of the classical Internet architecture, namely, the ping program \cite{KurRos-12, rfc1122}, which currently lacks a direct counterpart in quantum networks. In classical computer networks, the ping program serves as a straightforward tool to inquire about network-layer connectivity, enabling testing of node reachability and providing information on the latency involved.

Unfortunately, classical ping strategies are unable to fully test connectivity in quantum networks, which differs profoundly from classical network-layer connectivity due to entanglement. In quantum networks, connectivity arises from shared entangled states and is enabled by the capability to establish entangled links between two nodes, with the ``quality'' of these links related to the fidelity and robustness of the entanglement. 

To address this gap, we introduce the concept of \textit{Quantum Ping} (QPing), a tool designed to verify both the existence and the quality of entanglement links within a quantum network. Specifically, we design quantum primitives that form the foundation of diagnostic protocols, which can be deployed within different functionalities for the Quantum Internet \cite{Pirker_2019, Dahlberg_2019, Illiano_2022}. Indeed, tools that test entanglement-based connectivity are complementary to functionalities such as path selection, resource allocation, and quantum-specific foreseen Quality of Service (QoS) assessment for application requirements.

Our proposal is motivated not only by theoretical considerations but also by the growing need for scalable real-time diagnostics in emerging quantum network architectures \cite{Wei_2022, Azuma23}. We define several strategies for realizing the quantum ping, reflecting distinct operational and architectural assumptions. Active strategies aim to verify whether entanglement distribution can produce a useful entangled link, while passive strategies are designed to probe an entangled link from previously shared states. In each case, we make use of tools for entanglement verification such as sequential hypothesis testing \cite{Vargas2021, Pallister2018, Yu_2019, MiguelRamiro2022, Fields2024}, fidelity witnessing \cite{MiguelRamiro2023}, or Bell-type inequalities \cite{G_hne_2009}, and we analyze how time enters as a resource-limiting factor in the active cases. 

Time plays an important role not only in the execution of the protocol, but also in determining the required thresholds for entanglement quality. As quantum states degrade over time due to decoherence and operational errors, both the process of actively establishing remote entanglement and the quality thresholds themselves are affected. In addition, to assess the quality of entanglement, measurements are necessary, which in turn destroy part of the entangled quantum states. That is, the quality of a quantum connection cannot simply be assessed, but requires consuming some of the resources one actually would like to establish and use. While this is unavoidable, one clearly would like to minimize the required overheads.

Our approaches seek to optimize performance by avoiding resource-intensive methods such as quantum process tomography \cite{Mohseni2008, Cramer_2010}, which can be inefficient and provide excessive information beyond what is necessary. While both classical and quantum ping ultimately aim to answer a binary question, the quantum counterpart is far more challenging to resolve. In classical networks, the query is simply to check node-host reachability. In quantum networks, the question becomes whether the fidelity of the shared entanglement exceeds a (possibly time-dependent) threshold, a yes/no decision that is harder to answer because fidelity can degrade over time due to decoherence and operational errors. Moreover, quantum measurements tend to reveal more information than is strictly necessary to decide this binary outcome, making it difficult to design tests that are both resource-efficient and minimally invasive. This challenge is particularly relevant when assessing the feasibility of entanglement purification or distillation.

The paper is structured as follows. In Sec.~\ref{sec:back}, we review some basic concepts and related literature. In Sec.~\ref{sec:definition}, we properly define the concept of quantum ping and its underlying design framework. Furthermore, we introduce our system model by detailing the quantities of interest. In Sec.~\ref{sec:strategies}, we present different QPing strategies able to probe quantum connectivity in different network settings. Finally, we summarize our proposal and provide insights on further extensions of this work in Sec.~\ref{sec:conclusions}.

\section{Background}
\label{sec:back}
In this section, we briefly review some basic concepts and tools we make use of throughout this work, and we better substantiate the context of the quantum equivalent of ping. To this aim, it is useful to recall the main features of the classical ping program and its role in the classical Internet architecture.

\subsection{Classical Ping}
The classical ping utility is a fundamental tool in computer networks used to verify the reachability of a remote host through Internet Control Message Protocol (ICMP) messages \cite{rfc792}. 
In classical systems, the ICMP is used for error reporting and allows for assessing the nature of errors that occur at the network layer. The ping program sends ICMP messages to verify whether a destination is responsive. Its operation is very simple: a host sends an ICMP Echo Request to the IP address identifying the destination, and if no errors occur, that is, the network layer can successfully reach the destination, the node responds with an ICMP Echo reply \cite{KurRos-12, rfc1122}. 
This operation provides feedback about the network layer connectivity by testing the routing information \cite{KurRos-12, rfc1122}. Indeed, from the Echo reply, one can assert that the destination was reachable and evaluate an approximate round-trip time, giving a basic measure of latency.
As is known, the rationale behind the ping program stems from the best-effort nature of the network layer, which provides no guarantees regarding the ability to find a path to a destination. Clearly, the ping program does not guarantee reliable network layer services. In contrast, it tests network-layer connectivity by actively transmitting a packet, which triggers network-layer functionalities, thus testing the routing service. 

As a consequence, despite its simplicity, the ping program plays an essential role in real-time diagnostics, precisely because of its simple nature and binary outcome, i.e., “host reachable or not”. In this work, we explore how a conceptually analogous strategy could be defined in quantum networks, where connectivity involves fundamentally different resources and constraints. 

\subsection{Quantum Network Connectivity}
As mentioned above, in quantum networks, connectivity is no longer defined by the presence of a classical communication channel, but by the ability to establish quantum entanglement between two or more nodes. This quantum connectivity may be achieved either dynamically, by generating and distributing entangled states \cite{pompili2022}, or through local manipulation from pre-established resource states \cite{Pirker_2019}. 
However, several communication constraints arise, such as the impossibility of cloning unknown quantum information \cite{nielsen_chuang_2010}. Additionally, decoherence imposes temporal constraints, making quantum connectivity vanish over time, and the imperfect operations required to establish entangled links degrade the entangled resource. 
Given such constraints, it is essential to evaluate the \textit{quality} of entangled links through meaningful metrics to discern \textit{practically} \textit{useful} quantum connections. In this regard, several quantities may be considered, including the time required to establish the quantum connection, the noise model and its impact. 

In this paper, we refer to quantum connection quality as the similarity between the entangled links --entanglement shared between qubits held at remote nodes-- and the expected noiseless entangled state. More in detail, we consider fidelity serving as the primary metric and then further develop the tests and the quality metric according to different settings and QPing strategies.
This allows us to provide a comprehensive set of tools able to genuinely test quantum network connectivity for different network design architectures, as better described in the following.

\subsection{Network Architectures and Functionalities}
Quantum network architecture has been studied extensively, with several proposals aiming to organize their complex functionalities into layered stack models \cite{Pirker_2019, Dahlberg_2019, Illiano_2022}. These models typically include physical layers handling hardware and quantum memories, link layers responsible for entanglement distribution and purification, and network layers tasked with routing and resource management. However, the different proposed design approaches vary in their perspective and the design philosophy, ranging from bottom-up approaches \cite{Meter2013b,Gyongyo2017,Gyong2018,Pant2019} to top-down schemes \cite{Pirker_2018, Pirker_2019, Meignant2019, delocalizedinfo, Miguel_Ramiro_2023}. The former typically exploits the underlying physical topology of the network and aims to build entanglement connections along it. In contrast, top-down approaches circumvent physical constraints by leveraging pre-shared entanglement to create effective topologies on demand. These entanglement-based designs enable the dynamic construction of logical links through local operations and measurements. 

At the core of these architectures are entanglement distribution protocols \cite{Meignant2019, Navascues2020,Yu2021, Miguel_Ramiro_2023, Fan2024}. They include methods such as direct entanglement generation over physical channels, entanglement swapping via intermediate nodes (repeaters), and entanglement purification to enhance fidelity \cite{Briegel_Repeaters, D_r_2016, Meter_2013}. Efficient distribution is challenged by loss, noise, and memory coherence times, making reliable entanglement generation and distribution a non-trivial engineering task.

A subsequent key functionality is entanglement routing, which addresses the problem of identifying optimal paths through a network to establish entanglement between distant nodes \cite{Lee_2022,Shi_2024,Abane_2025}. Routing protocols must balance factors like path fidelity, resource availability, and latency, also relying on classical signaling.

Quantum ping protocols should be designed to suit the requirements and structural features of each network architecture approach. To this end, we present a general formulation of tests and quality metrics that can be applied across different architectures. While the formulation does not require a detailed knowledge of the specific entanglement distribution or routing mechanisms, it assumes that the core functionalities for entanglement generation, distribution, and routing functionalities are available and can serve the quantum ping operation.


Specifically, in its active variants, the quantum ping strategies trigger generation and distribution of entanglement over a candidate path, implicitly testing both connectivity and management. Furthermore, the segment-based variant allows to test quantum connectivity at link-level, providing information on single-hop connectivity. Finally, the passive strategy leverages pre-shared entangled resources and exploits functionalities such as routing based on resource state and entanglement control to extract an entangled state through local operations.

In conclusion, the proposed QPing strategies inherently exploit entanglement generation and distribution, as well as entanglement-aware routing functionalities. Consequently, a successful QPing execution implicitly verifies the correctness, consistency, and availability of the underlying routing data. In particular, active QPing strategies directly test routing information by confirming the physical availability and viability of the selected entanglement distribution path. In contrast, passive QPing approaches validate the logical link structure resulting, ensuring that locally manipulated entanglement resources indeed match the routing topology.

To the best of our knowledge, such a concrete formulation of a quantum ping primitive has not been addressed so far.

\section{Quantum Ping Design} 
\label{sec:definition}

\subsection{Design Objectives}
Given the above discussion, a quantum ping test for quantum connectivity should answer the question ``Is there usable entanglement?'', where usable entanglement refers to entangled states that pass a specified quality criterion. The design of quantum ping strategies is operatively translated into the definition of such a quality criterion and its test. Towards this design, we aim to preserve the features that make the classical ping effective, namely, lightweight network resource consumption and simplicity. 
Notably, these two features are particularly challenging to achieve in the context of quantum networks. Indeed, to observe the quality of an entangled state, this should be measured and hence consumed. 

In this regard, the strategies for quantum state reconstruction such as quantum state tomography and quantum process tomography, which respectively characterize quantum states and quantum channels, are heavily resource-consuming and difficult to scale as the number of parameters to be determined exponentially increases with the dimension of the space of the system \cite{Mohseni2008, Cramer_2010}. As a consequence, although tomography allows for the characterization of an entangled link, it is not a viable strategy due to complexity and resource overhead.

To mirror the simple nature of classical ping, we avoid tomography and aim to design minimal tests for quantum connectivity under temporal and hardware limits: finite memory and imperfect operations constrain purification attempts and maximum fidelity. 
Consequently, the possibility of entanglement purification and fidelity thresholds depends on each setting capabilities, so QPing procedures must be tailored to each scenario.

\subsection{System Model}
We consider a quantum network consisting of nodes, connected by quantum channels, that is capable of distributing entanglement. Nodes may function as users, repeaters, or routers, and the network may operate under various considerations. 

\begin{figure}[t!]
    \centering
    \includegraphics[width=0.85\linewidth]{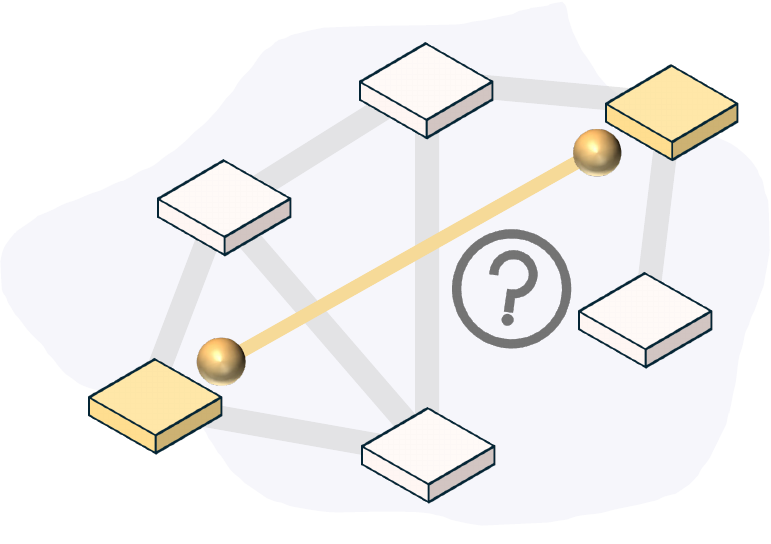}
    \caption{General idea of Quantum Ping (QPing). Two nodes of the network aim to decide whether they can generate and share useful entanglement between them.}
    \label{fig:Ping_general}
    \hrulefill
\end{figure}

We explore different scenarios and possible definitions of a quantum ping, grounded in the unique characteristics of quantum networks. The overall underlying goal is to check whether a quantum connection can be established between two parties, that is, whether useful entanglement can be generated between them, and, in some cases, to witness its quality (see Fig.~\ref{fig:Ping_general}).

Specifically, given two quantum nodes and a quantum task or application $T$ (e.g., purification, teleportation, QKD, secret sharing), we can formally define the minimal required fidelity.

\begin{definition}[\textit{Minimal required fidelity}]
Let $A$ and $B$ be two network nodes sharing a quantum connection and let $T$ be a quantum task or application. The minimal required fidelity $F_0^T$ for task $T$ is the smallest fidelity such that $T$ can be successfully performed. Formally:
\begin{equation}
    F(\rho_{AB}, \ket{\Phi^+}) = \bra{\Phi^+} \rho_{AB} \ket{\Phi^+} \ge F_0^T,
\end{equation}
where $\rho_{AB}$ is a shared bipartite quantum state required to perform $T$, and $\ket{\Phi^+} = \tfrac{1}{\sqrt{2}}(\ket{00} + \ket{11})$ is the reference Bell state.
\end{definition}

\noindent Note that this definition can be extended to more general kinds of entangled, including multipartite or higher dimensional, states.

Different tasks may then impose different \(F_{0}^{T}\), so the ping procedure is tailored to ensure $\Pr\bigl(F(\rho_{AB}, \ket{\Phi^+}) \ge F_{0}^{T}\bigr)$ exceeds some desired confidence level, specified according to the QPing strategy. 

This formulation ensures that QPing answers the precise question: “Is there usable entanglement between \(A\) and \(B\) for application \(T\) under the given temporal and hardware limitations?”
%

\begin{quantumping}
    Given access to a quantum network, determine whether two nodes $A$ and $B$ can establish an entangled state $\rho_{AB}$ such that $F(\rho_{AB}, \Phi^+) (t)>F_{0} (t)$ with high confidence, and using minimal quantum resources.
\end{quantumping}

Observe how, in analogy with the classical ping, the time to establish a quantum connection should be incorporated as a central metric in the protocol. However, rather than referring to the runtime of the ping process itself, we focus on the expected time required to generate a usable quantum connection, that is, to successfully distribute entanglement between two end-nodes. This time is not only an observable output of the protocol, but also a factor that influences the threshold parameter $F_0$, which determines whether the resulting entanglement is useful for further applications. Specifically, longer distribution times may degrade $F_0$, due to the finite coherence time of quantum memories and the accumulation of operational errors. 

For example, if a QKD application requires \(F_{\rm req} = 0.9\), but operational errors and memory decoherence during the process of establishing entanglement causes the fidelity to drop to \(0.85\) within \(100\,\mathrm{ns}\), then the state must be initially prepared with \(F_{\rm init} > 0.95\) to ensure it still satisfies \(F(t_{\rm use}) \geq F_{\rm req}\) at the time of use. Therefore, QPing strategies must internalize this time-to-use notion, rather than a simplistic “distribution time”, to remain compatible with the temporal and operational constraints of quantum hardware.

Different effects can affect the dynamical definition of $F_0(t)$, among which we can highlight:

\begin{itemize}
    \item \textit{Memory decoherence times.--} Note that the entanglement quality decays with memory decoherence time, generally speaking as $F_0(t) \approx F_0 \exp(-t / \tau)$. Classical messaging and delays from classical processing would contribute to such decoherence.
    
    \item \textit{Gate and operational errors.--} Entanglement swapping operation between not perfect states leads to lower fidelity resulting states. 
    Imperfect implementations of swapping operations also accumulates noise, leading to reduced fidelities $F_0(t) \approx F_0 \exp(-t / \tau)  q_{gate}^{N_{gates}} $.
    
    \item \textit{Entanglement distribution times.--} Entanglement swapping operations may be non deterministic, leading to stochastic distribution times that depend on the number of attempts required for success. In general, $F_0(t) \approx F_0 \exp(-t_{dist} / \tau)$, where $t_{dist}$ is the average distribution time that depends on the attempt success probability, per trial $\approx 1/p$.
\end{itemize}
Note that these effects not only impact the threshold fidelity required for an application, but also limit the maximum achievable fidelities. If, for instance, decoherence and gate errors reduce any initially prepared state below the application threshold within the usable time window, no protocol will succeed. Therefore, QPing must not only evaluate whether a state meets \(F_0\), but also whether the network can realistically produce such a state under its operational constraints, intrinsic to the network hardware and architecture.

As we describe in the scenarios below, the evaluation of this time metric will differ depending on the protocol model, whether based on segments, end-to-end connections, or pre-established resource states. We refer to Table~\ref{tab:ping_comparison} for a comprehensive summary and comparison between the strategies we introduce. 

We note that the specific solutions presented here are intended as proof-of-concept examples. In practice, more sophisticated or alternative tools could be employed depending on the network architecture and application requirements.

\section{Quantum Ping Versions and Operational Strategies}
\label{sec:strategies}
In the following, we detail the proposed strategies, which can be classified into active and passive categories. Among the active strategies, we distinguish between those aimed at evaluating an entire path --provided by the routing functionality-- and those focused on testing individual segments, namely, fractions of a given path. Finally, we consider passive strategies for testing entangled links obtained from a pre-shared resource.

\begin{remark}
    Throughout this work, we refer to a QPing strategy as \textit{active} when it explicitly triggers entanglement distribution processes across the network. Conversely, a strategy is labeled as \textit{passive} when it operates on pre-established resources, such as those resulting from background distribution protocols already executed, entanglement reservoirs, or multipartite resource states \cite{Dahlberg_2019,DaiRinTow-21,Freund2024}. 
    
    This terminology differs from the conventional terminology in classical networking, where active routing refers to the dynamic computation of paths during data transmission, as opposed to session routing, where a route is fixed during virtual circuit setup and followed passively thereafter \cite{Tan-10}. In both cases, classical routing decisions are concerned with the path, whereas in QPing, our distinction is orthogonal and focuses on whether the diagnostic process activates the entanglement generation and distribution process or probes existing entanglement. This distinction is subtle but crucial, as it accommodates both proactive and reactive entanglement routing paradigms, where routing decisions and entanglement distribution occur in varying orders \cite{Abane_2025}. 

    From the perspective discussed in \cite{Abane_2025}, QPing active strategies \textit{stimulate} both internal and external phases of routing functionality, forcing the network to instantiate the logical route physically through entanglement distribution. Differently, passive strategies probe the existing state and prior routing decisions, intended as pre-selected entangled links and a sequence of Pauli measurements. 
\end{remark}

While the proposed QPing primitive is conceptually aligned with the output of the classical ping, its implementation and operational semantics depart significantly due to the fundamental characteristics of quantum networks. Specifically, and as remarked, it is necessarily interleaved with key quantum network functionalities such as entanglement generation, distribution, routing, and management. As these functionalities do not have a direct mapping in the classical Internet protocol stack \cite{Illiano_2022}, and the design of abstract models for quantum networks is ongoing, we avoid the rigorous definition of diagnostic protocols for quantum networks. In light of this, the proposed QPing is not coupled to a specific protocol stack or architectural models. Differently, it can be conceived as a \textit{functionality-driven} primitive, that is, its applicability depends on the availability of core functionalities and abstracts from their logical structure and implementation. As a consequence, our proposal can be further deployed within centralized architectures, where the above-mentioned key functionalities are delegated to a single managing network entity. Besides, it provides the quantum-aware preliminary tests that can be leveraged within more complex distributed approaches.

As a consequence, our proposal represents a layer-agnostic module that serves as a \textit{primitive} for diagnostic protocols in quantum networks.

\subsection{Path-based Active QPing}
\label{sec:active_path}
\begin{figure}
    \centering
    \begin{subfigure}[b]{0.38\textwidth}
        \centering
        \includegraphics[width=1\textwidth]{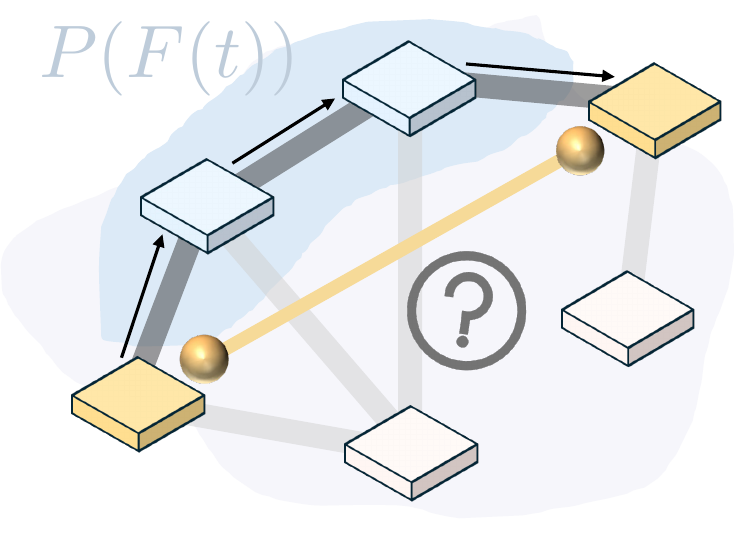}
        \caption{Path-based active QPing protocol.}
        \label{fig:activeA}
    \end{subfigure}
    
    \vspace{0.5cm}
    \begin{subfigure}[b]{0.38\textwidth}
        \centering
        \includegraphics[width=1\textwidth]{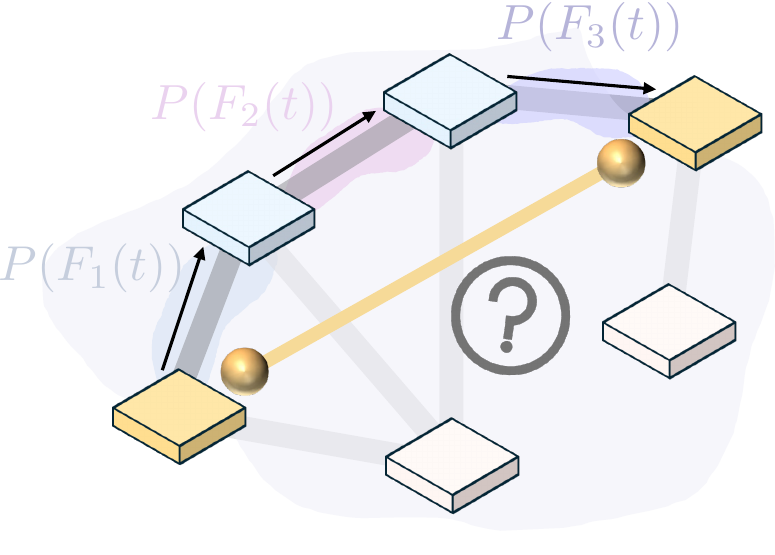}
        \caption{Segment-based active QPing protocol.}
        \label{fig:activeB}
    \end{subfigure}
    \caption{Active Quantum Ping strategies. (a) Path based. A whole path between the end nodes is tested through local strategy, where measurements are performed at both end nodes, or global strategy, where one qubit is bounced back through the path and global measurement is performed at one end node. (b) Segment based. The individual segments composing the path between the end nodes are simultaneously and individually tested.}
    \label{fig:active}
    \hrulefill
\end{figure}

The first scenario is pictorially represented in Fig~\ref{fig:active} and referred to as \textit{path-based} active ping. This QPing version enables two end nodes to test whether a useful quantum link can be established through a given path, selected by the network routing process or, if the path computation has not been performed, computed by a path finding algorithm. In this scenario, the path-based QPing aims at answering the question ``Is the fidelity of the entanglement that can be generated between these two nodes high enough to perform a certain task?''. More in detail, we aim to assess whether the fidelity of the entangled state distributed over the path exceeds the threshold fidelity. 
As discussed before, this fidelity threshold will depend on factors such as the type of entanglement (e.g., bipartite or multipartite), decoherence, memory decay effects, and the operational noise (e.g., the quality of local gates). These variables define a threshold fidelity, $F'_0 (t)$, typically lower than the one required for the application itself, and the problem ultimately reduces to a fidelity witnessing problem \cite{MiguelRamiro2023}.

If the end nodes are capable of storing multiple entangled pairs and performing entanglement distillation prior to application, then the QPing diagnostic question becomes: “Is the fidelity of the entanglement that can be generated between these two nodes high enough to allow successful purification?” Importantly, the availability of entanglement distillation does not alter the structure of the QPing protocol itself, but rather only affects its operational parameters.

The fidelity witnessing problem consists in deciding whether the fidelity of a quantum state lies above or below a given threshold --typically framed as a hypothesis testing task-- while incurring minimal overhead. In Ref. \cite{MiguelRamiro2023}, we proposed different strategies to tackle this problem using only local operations. While not optimal, since they reveal slightly more information than strictly needed, they offer efficient solutions directly applicable to address the QPing problem.

In our proposal, we resort to sequential hypothesis testing \cite{Vargas2021, Pallister2018, Yu_2019, MiguelRamiro2022, Fields2024}, following the strategies we proposed in Ref. \cite{MiguelRamiro2023}, integrating both local (end-node) and global (bouncing) strategies. The protocol gathers only the essential information required to solve a binary problem with high probability and confidence, dynamically adapting the solution over time. This approach ensures efficient information collection while balancing accuracy and speed, with continuous adaptation to evolving data and time-dependent factors, making it ideal for real-time quantum systems.

In the context of QPing, however, one can consider a variant in which a Bell pair is prepared at one end, send one qubit through the path to the other node, and then send it back so that the original node holds both particles again. A single joint (global) Bell-state measurement can then estimate the fidelity of the full round-trip. For this reason, this variant is also referred to as global strategy. This uses fewer local operations but doubles the noise (each link is traversed twice). While resource-saving, the double passage can importantly reduce fidelity and even break entanglement; hence, one must carefully adjust the threshold to account for this. In practice, this bouncing strategy might be useful for quicker checks or for individual segment testing (see below), but it carries the risk of misleading results if the path is very noisy.

In the following, we summarize the two variants of the path-based active QPing with a step-by-step description, which includes dynamical time evaluation and is based on statistical decisions (see also Fig.~\ref{fig:active}).

\medskip 

\subsubsection{Active End-node Strategy}

\begin{enumerate}[label=(\roman*)]

    \item \textit{Path assumption.--} Given a path between two end nodes, choose a prior distribution $P(F(t))$ over fidelity values, reflecting prior path knowledge. If no such knowledge exists, a uniform prior may be assumed. 

    \item \textit{Collect data.--} Establish entangled states between end nodes and perform local measurements of random Pauli correlations \cite{MiguelRamiro2023}. After $n$ trials, let $k$ be the number of ``passes'' observed at time $t$. By Hoeffding/Chernoff-type bounds, the frequency $k/n$ concentrates around its expectation, providing statistical confidence \cite{MiguelRamiro2023}. This process reveals the pass/fail counts needed for fidelity witnessing.
    One can tune and actively update the protocol efficiency as follows.

    \item \textit{Statistical update.--}   Compute the posterior distribution $P(F \mid \text{data})$ using Bayes theorem:
   \begin{equation}
   \label{eq:Bayes}
    P(F(t) \mid k,n) = \frac{P(k \mid F(t), n) \cdot P(F(t))}{\int_0^1 P(k \mid F(t), n) \cdot P(F(t)) \, dF},
   \end{equation}
   where $p(F(t))$ models the probability of a ``pass'' given fidelity $F(t)$.

    \item \textit{Compute Posterior Probability.--} Evaluate:
   \begin{equation}
   \mathbb{P}(F(t) \geq F_0(t) \mid \text{data}) = \int_{F_0(t)}^1 P(F(t) \mid \text{data}) \, dF.
   \end{equation}

   \item \textit{Decision rule (hypothesis test).--} Choose a confidence threshold $\eta \in [0,1]$, and an uncertainty parameter $\delta$. Then:
   
\begin{itemize}
  
    \item[--] If $\mathbb{P}(F(t) \geq F_0(t) \mid \text{data}) \geq \eta$, ``accept'' the link (ping success). Otherwise ``reject'' the link (ping fail).
    
   
    \item[--]  Optionally: if $\mathbb{P}(F(t) \geq F_0(t)) \in [\eta-\delta, \eta+\delta]$, declare ``inconclusive''.
       
\end{itemize}
\end{enumerate}

At each step, time advances (more pairs are tested), and the posterior probability distribution is updated adaptively. This sequential procedure naturally balances resource use and decision confidence: as soon as sufficient evidence accumulates, one reaches a decision without wasting trials. This sequential QPing minimizes the expected number of entangled-pair trials to reach a reliable accept/reject verdict. Explicit cost functions (e.g. expected number of pairs used as a function of true $F$ and thresholds) could be derived using standard sequential hypothesis testing results \cite{Vargas2021, Pallister2018, Yu_2019, MiguelRamiro2022, Fields2024}, but such more practical analysis falls outside the conceptual scope of this work. 

Importantly, notice how a successful QPing diagnosis indirectly confirms the presence of an operational routing path between the end nodes, thereby validating both physical connectivity and entanglement feasibility.

\begin{remark}
    This strategy suits well \textit{heralded swapping} policies, i.e., policies where routers carry out swapping operations based on the swapping outcomes of other routers \cite{Abane_2025}. Since heralding already introduces sequential decision points, the path-based strategy can run alongside to confirm that the entire intended route is still entangled before committing to the task operations.
\end{remark}

\subsubsection{Active Bouncing Strategy}

\begin{enumerate}[label=(\roman*)]

    \item \textit{Path assumption.--} Given a path between two end nodes, choose a prior distribution $P(F(t))$ over fidelity values of the \textit{two-way} path.

    \item \textit{Collect data.--} A Bell pair is prepared locally and one half is sent through the quantum path and reflected back. Perform global operations (i.e., projection onto $|\Phi^+\rangle$) to gather statistics more efficiently. 

    \item \textit{Statistical update.--} From the data (e.g., \(k\) successful outcomes in \(n\) trials), update the posterior distribution over the round-trip fidelity \(F_{\text{rt}}(t)\), defined as the effective fidelity after the qubit traverses the path twice. Since the protocol performs a round-trip transmission and measures global Bell-state projections, the probability of success per trial is given by the effective round-trip fidelity. For depolarizing channels it equals to: \begin{equation}
    P_{\text rt}(F) = F^2 + \frac{(1 - F)^2}{3}.
    \end{equation}
    Then, using Bayes theorem Eq. \eqref{eq:Bayes}, the posterior distribution  $P(F_{\text rt}(t) \mid k, n)$ is computed.

    \item \textit{Compute Posterior Probability.--} To test whether the original one-way channel is suitable, compare the inferred round-trip fidelity against the original target threshold \(F_0\). Specifically, compute:
    \begin{equation*}
    \mathbb{P}(F_{\rm rt}(t) \geq F^{\rm eff}_0 (t) \mid \text{data}) = \int_{F^{\rm eff}_0(t)}^1 P(F_{\rm rt}(t) \mid \text{data}) \, dF,
    \end{equation*}
    where ${F^{\rm eff}_0(t)}$ represents the effective threshold taking into account the double use of the channel.

   \item \textit{Decision rule (hypothesis test).--} Choose a confidence threshold $\eta \in [0,1]$.
   \begin{itemize}
      \item[--] If $\mathbb{P}(F_{\rm rt}(t) \geq F_0^{\rm eff}(t)) \geq \eta$, accept the link (ping success). Otherwise, reject the link (ping fail).

   \end{itemize}
   
\end{enumerate}
This bouncing strategy provides more powerful information per trial, but at the cost of increased noise as the path is used twice. 

\begin{prop}[\textit{Bouncing noise factor}]
In the bouncing (global) QPing variant, a single qubit traverses each link twice. If each one‑way channel is e.g. a depolarizing map of fidelity \(f\), then the round‑trip channel has fidelity
\begin{equation}
F_{\rm rt} \;\geq\; F^2,
\end{equation}
which for a depolarizing channel equals $F_{\rm rt} \;=\; F^2 + \frac{(1-F)^2}{3}$.
\end{prop}

Consequently, to test against a target threshold \(F_0\), one must use an \emph{effective} threshold of
\(\;F_0^{\rm eff} \geq {F_0}^2\)
when interpreting bounce‑based measurements (e.g., equaling $F_0^{\rm eff} = F_0^2 + \frac{(1 - F_0)^2}{3}$ in the depolarizing case), up to time decay adjustments.

In both settings, QPing can be evaluated at different times, where the threshold fidelity is dynamically taken into account, considering the features discussed above. Notably, prior knowledge of $P(F(t))$ can also enable more efficient strategies and help determine whether it is worthwhile to run the QPing protocol, as detailed in the following proposition.

\begin{prop}[\textit{Prior QPing decision}]
Let \(P(F)\) be the prior \textit{known} distribution over end‑to‑end fidelity \(F\), and let \(F_0(t)\) be the current threshold at time \(t\).  Define
\begin{equation}
P_{\rm 1} \;=\;\int_{F_0(t)}^1 P(F)\,dF,
\quad
P_{\rm 2} \;=\; 1 - P_{\rm 1}.
\end{equation}
Given a confidence level \(\eta\in(0,1)\):
\begin{itemize}

  \item[--] If $P_{1} \gg \eta$, the link is almost certainly usable (skip QPing).
  
  \item[--] If $P_{2} \gg \eta$, the link is almost certainly unusable (abort QPing).
  
  \item[--] Otherwise, \(P_{\rm 2}<\eta<P_{\rm 1}\), the outcome is unclear (run QPing to resolve feasibility).
\end{itemize}
\end{prop}

In both the end-node and bouncing versions of the active quantum ping,  local twirling techniques  \cite{StandForms, RandoCompiling} may be applied to simplify noise models. In the global strategy, random Clifford twirling is feasible and transforms the effective channel into a depolarizing form, allowing for efficient fidelity witnessing via Bell measurements. In the end-node case, local Pauli twirling yields a Bell-diagonal state, simplifying the structure of the fidelity witnessing and improving statistical robustness, although full Clifford symmetrization is not implementable with local operations alone.

Remarkably, if the end nodes are able to store multiple entangled copies and perform collective operations on them, the efficiency of hypothesis testing can be exponentially improved \cite{MiguelRamiro2022, MiguelRamiro2023}. This enables an optimized strategy, in both local and global strategies, for gathering the required statistics in the QPing protocol.

\subsection{Segment-based Active QPing} \label{sec:active_segment}
The path-based QPing requires establishing full end-to-end connections solely to check the viability of a quantum link. However, in scenarios of highly loaded networks, this can be too demanding in terms of resources. We therefore explore an alternative strategy based on testing the individual segments that form the path between the end nodes, see Fig.~\ref{fig:active}~(b), based on the same strategies introduced above, i.e., making use of the end-node and bouncing approaches. The segment-based strategy tests individual path segments, either identified through the network routing functionality or determined by a link-level discovery process.

\begin{prop}[\textit{Segment‐based composition}]\label{thm:segment_composition}
Consider a path of $N$ segments, where segment $i$ produces a Bell pair of fidelity $F_i$ (depolarizing form) and each entanglement‐swapping operation introduces a fixed fidelity factor $q_{\rm swap}$. Under independent depolarizing noise on each link and ideal LOCC at the end nodes, the resulting end‐to‐end fidelity satisfies:
\begin{equation}
    F_{\rm end} \;=\;\Bigl(\prod_{i=1}^N F_i\Bigr)\; \;q_{\rm swap}^{\,N-1},
\end{equation}
up to time decay effects. 
\end{prop}

This method offers clear advantages in terms of reduced overhead and parallelizability, as all segments can be tested simultaneously. These features make the bouncing-based approach particularly viable in scenarios requiring rapid, parallel diagnostics over channels with relatively low noise. Its simplicity also facilitates deployment in near-term architectures. The steps for the segment-based ping strategy largely mirror those of the path-based ping, but are applied independently to each segment. 

In cases where entanglement purification is available within individual segments, as in standard repeater settings  \cite{Briegel_Repeaters, D_r_2016, Meter_2013}, a key advantage is that the ping decision about the viability of establishing entanglement is not underestimated: it suffices to meet a threshold fidelity locally within each segment, rather than across the full end-to-end connection. This approach allows for a more modular and realistic assessment in various scenarios, such as high-loaded networks --where resource reservation for end-to-end connections might be strictly limited in time-- or in cases where the path involves networks managed by different controllers. Although no global path is used in this strategy, successful certification across all segments effectively confirms that a route can be assembled via entanglement swapping.

\begin{remark}
    The segment-based strategy suits unheralded swapping strategies, where each router performs the swapping independently and in parallel without awaiting the swapping outcomes of the others. The segment-based strategy can identify which segments are failing; furthermore, the results are useful for offline diagnostics to improve resource allocation in the next rounds.
\end{remark}

However, this segmented approach might not capture the true performance of the entire end-to-end quantum channel. In particular, it neglects critical inter-node operations such as entanglement swapping, which may introduce nontrivial noise and failure that are invisible when testing links in isolation. To address this, length-two segments can be tested, including entanglement swapping operations at indeterminate nodes. This allows to determine the achievable fidelity for each segment, therefore deciding whether an entangled link, with or without entanglement purification at intermediate nodes, is viable. 

This independence complicates the treatment of time, particularly when attempting to reconstruct the end-to-end entanglement generation performance. Since no actual end-to-end quantum state is created in this approach, one must instead extrapolate from individual segment metrics, while accounting for the expected behavior of entanglement swapping operations at intermediate nodes. Moreover, in case the bouncing approach is implemented, non-Markovian effects on the channel noise when bouncing back the particles could arise and affect the correct evaluation of the round-trip QPing.  Importantly, additional delays introduced by swapping, along with potential decay in neighboring links that are ready earlier, must also be considered.

Both active QPing strategies can be combined and implemented on demand, suggesting that QPing should not be limited to a single test. Instead, it may consist of a suite of connectivity checks tailored to the network topology and capabilities. For instance, an end-to-end test based on entanglement distribution could be complemented by a segment-based test relying on local entanglement purification, both forming part of a more general and flexible QPing protocol.

\begin{figure}[t!]
    \centering
    \includegraphics[width=0.85\linewidth]{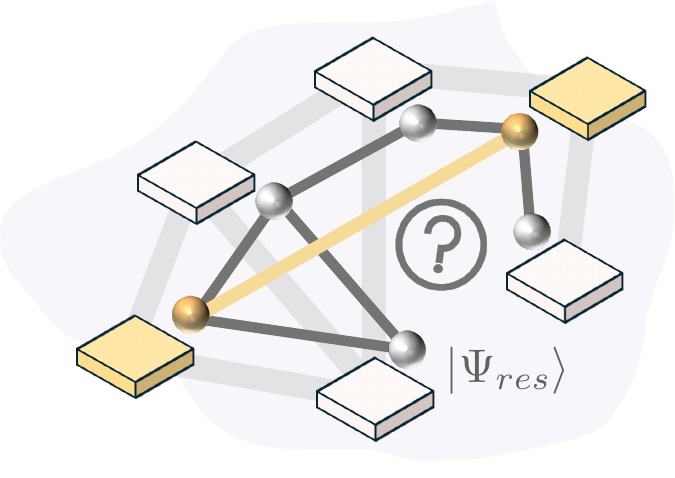}
    \caption{Passive quantum ping. Entanglement is already pre-shared between network constituents, and the ping protocol decides whether suitable (local) manipulation of the resource state can lead to direct and useful entanglement between end nodes. }
    \label{fig:passive}
    \hrulefill
\end{figure}

\subsection{Passive QPing}
The last scenario considers the case where a shared entangled resource state has already been established among the network nodes, i.e., an entanglement-based quantum network \cite{Pirker_2018, Pirker_2019}, and the end-to-end connection is achieved via local operations on this resource. In this situation, the task consists of determining, first, whether the two end nodes are part of the same entangled state, and second, whether they can be suitably connected through appropriate local manipulations, see Fig.~\ref{fig:passive}.

Because identifying entanglement between specific parties within a large multipartite state is generally challenging \cite{Dahlberg2018, Dahlberg2020} without multipartite state characterization, we assume prior knowledge of the structure of the entangled resource $|\Psi_{\text{res}}\rangle$ and the protocol that ideally transforms it into the desired configuration. In particular, we consider that the network consistently generates a known entangled state during idle periods, allowing a predefined LOCC (local operations and classical communication) protocol to be used for establishing bipartite connections between selected nodes. For instance, if the resource state consists of a 1D cluster state, just applying Pauli Y measurements in the intermediate qubits and Pauli Z measurements on other neighbors achieves the desired end-to-end connection \cite{Hein2004, Hein2006}. Similar properties exist for more complex resource state structures, such as 2D cluster states \cite{Freund2024, MorRuiz2024, MorRuiz2025}.

Here, the fidelity ceases to be the most relevant figure of merit, since the goal is no longer to estimate how close a state is to a given target to, for instance, perform entanglement distillation, but rather to determine whether entanglement can be actually present. We therefore shift our focus to a more suitable quantity in this context: entanglement witnessing, although similar tools as before can be employed. The goal is to discern whether entanglement can be extracted between the end nodes, independently of local unitaries, for practical use. We can define passive quantum ping as follows.

\begin{table*}[htbp]
\centering
\caption{Comparison of the different quantum ping strategies proposed}
\label{tab:ping_comparison}
\renewcommand{\arraystretch}{1.2}
\begin{tabular}{@{}p{5cm}p{8cm}@{}}
\toprule
\multicolumn{2}{l}{\textbf{Active Path-Based}} \\ \midrule
Time Sensitivity         & Yes \\
Operations               & Local or global (end-node or bouncing strategy) \\
Certification Method     & Sequential hypothesis testing \\
Information per Trial    & Medium (requires multiple rounds for confidence) \\
Robustness               & Medium (whole path may be too sensitive to noise) \\

\end{tabular}

\vspace{0.2cm}

\begin{tabular}{@{}p{5cm}p{8cm}@{}}
\toprule
\multicolumn{2}{l}{\textbf{Active Segment-Based}} \\ \midrule
Time Sensitivity         & Partial (time per segment small, but global info indirect) \\
Operations               & Local or global (and simultaneously for each segment) \\
Certification Method     & Sequential hypothesis testing \\
Information per Trial    & Low-medium (not directly end-to-end) \\
Robustness               & High (no dependence on swapping or coordination) \\
\end{tabular}

\vspace{0.2cm}

\begin{tabular}{@{}p{5cm}p{8cm}@{}}
\toprule
\multicolumn{2}{l}{\textbf{Passive Resource-Based}} \\ \midrule
Time Sensitivity         & Low (resource state pre-shared, conversion is quasi-immediate) \\
Operations               & Local \\
Certification Method     & Entanglement witnessing, Bell inequality \\
Information per Trial    & High (if resource is high-quality, few samples suffice) \\
Robustness               & Low (require in general high quality entanglement) \\
\end{tabular}

\end{table*}

\begin{passivequantumping}
    Determine whether two target nodes $A$ and $B$, assumed to be part of a pre-shared entangled resource state $|\Psi_{\text{res}}\rangle$, can be locally transformed (for a given fixed LOCC) into a high-quality entangled state $\rho_{AB}$, and possibly verify the entanglement of $\rho_{AB}$.
\end{passivequantumping}

An important distinction in this scenario is that time evaluation is not as crucial, since high-quality entanglement is assumed to be pre-established during idle network functioning \cite{Pirker_2018, Pirker_2019}. Nodes may generate and store entangled links continuously, so that when a ping is issued, the entanglement is already available. This greatly reduces latency. Only local manipulation of the resource state is needed, which can be treated as quasi-timeless, although certain time considerations could need to be taken into account \cite{MorRuiz2025}. This can significantly reduce network idle times compared to schemes requiring repeated entanglement generation and coordination. 

\begin{prop}[\textit{Passive QPing criterion}]\label{prop:passive_structural}
Let $|\Psi_{\rm res}\rangle$ be a pre‐shared graph‐state resource on a network graph $G=(V,E)$. Two nodes $A,B\in V$ can extract a Bell pair by fixed LOCC if and only if there exists a path $P$ from $A$ to $B$ in $G$. Equivalently, the localizable entanglement between $A$ and $B$ is nonzero precisely when $G$ contains a connecting path.
\end{prop}

The goal is then to determine whether two end nodes can extract, via local operations, a usable bipartite entangled state. This approach does not rely on active path selection or channel establishment and is essentially unaffected by temporal constraints. After performing the local transformation to distill the bipartite state $\rho_{AB}$, the protocol verifies its entanglement either via fidelity witnessing \cite{Miguel_Ramiro_2023}, Bell-type inequalities violation \cite{Bartkiewicz2013, Friis2018}, or entanglement witnesses \cite{Brandao2005, Hyllus2005, Eisert2020}, depending on the expected quality of the output. Since almost no dynamical process is involved, the success criterion becomes less time-independent and mainly structural. 

The steps for the passive quantum ping strategy can be summarized as follows.

\subsubsection*{Passive Quantum Ping}

\begin{enumerate}[label=(\roman*)]
    \item \textit{Local procedure.--} Apply the known LOCC procedure to transform $|\Psi_{\text{res}}\rangle$ into a bipartite state between $A$ and $B$.

    \item \textit{Verify entanglement.--} Choose a witnessing strategy:
    
    \begin{itemize}
    \item[--] If $\rho_{AB}$ is expected to be high-quality (either present or not): use binary entanglement witness or a Bell inequality test.

    \item[--] If possible low-quality entanglement is expected (e.g., some edges were faulty or missing):  use fidelity witnessing based on sequential hypothesis testing as in the active ping scenario.
    \end{itemize}

    \item \textit{Collect data for the chosen witness strategy}. 

    \item \textit{Decision rule.--} Given the chosen witnessing strategy (up to confidence):
    \begin{itemize}
    \item[--] If using entanglement witness  $W$: Check whether it satisfies $\text{Tr}[W \rho_{AB}] < 0$.
    
     \item[--]If using CHSH-like inequality: $|S| > 2$ implies entanglement.

    \item[--] If using fidelity witnessing: Accept if ${F} \geq F_0$. 
    \end{itemize}

\end{enumerate}

The choice of certification strategy depends strongly on the network setting, on the available prior knowledge, and on the desired confidence level for application. Bell inequality tests offer a device-independent route, but demand high experimental precision and a strong violation margin, typically requiring high-fidelity entangled states and detailed knowledge of the target expected states. Entanglement witnesses, on the other hand, can be very efficient in specific scenarios, particularly when the expected entangled state is well known, but it can also be partially state dependent. They can however fail to guarantee that there is no entanglement. Fidelity witnessing provides a balanced alternative: while not device-independent, it enables efficient, quantitative certification of entanglement with modest experimental overhead and is robust under noise, although it assumes knowledge of the target state.

\begin{remark}
    With reference to the taxonomy provided in \cite{Abane_2025}, the passive strategy naturally couples with proactive or virtual routing, where the external phase, i.e., generation of end-to-end connections, follows the path computation and leverages already established entangled links.
\end{remark}

\section{Outlook and conclusions} \label{sec:conclusions}
We have introduced the concept of a \textit{Quantum Ping} (QPing) protocol as a diagnostic primitive tool for quantum networks. Drawing inspiration from its classical analogue, the QPing aims to probe connectivity between two quantum network nodes, understood here as the capacity to establish useful entanglement, using minimal resources and with operational relevance for future quantum networks. This tool can complement entanglement routing by providing dynamic and efficient connectivity information to support path discovery and routing decisions, thereby enhancing quantum network robustness and scalability. From this perspective, QPing can be seen as a pre-screening tool for quantum network Quality of Service (QoS): by comparing current network performance indicators against time-dependent thresholds, one can decide whether attempting entanglement is worthwhile or whether performance upgrades are required.

We propose several realizations of the quantum ping protocol, each considering different architectural constraints and resource assumptions. These include (i) an \textit{active path-based} strategy where entanglement is established dynamically across an entire path and verified using local or global witnessing methods; (ii) a \textit{active segment-based} variant where individual segments are benchmarked simultaneously and independently; and (iii) a \textit{passive strategy} in which pre-shared entangled resource states are manipulated locally to extract and certify bipartite entanglement between the target nodes. 
While each of these scenarios is meant for particular settings, realistic networks may require hybrid schemes. For example, if two nodes are not directly connected in a graph-state resource, one might perform local measurements to route entanglement through intermediate nodes. Exploring such semi-active pings, is a promising direction for future work.

Further directions also include studying quantum ping in dynamic and realistic settings --where topology or noise levels change over time-- and extending the framework to multipartite scenarios, while integrating the tool within quantum networks stack layer structures. Beyond specific implementations, the QPing abstraction provides a conceptual bridge between lower-layer entanglement-based protocol performance and higher-layer utilization and control, enabling resource-efficient entanglement diagnostics.

In a similar direction, integrating QPing into a standard sequential hypothesis‐testing framework would offer several benefits: it would enable rigorous control of error rates, yield explicit bounds on the required number of trials, and facilitate incorporation of well‐established results from the classical sequential‐analysis literature.

Additionally, one could also consider ping strategies in device-independent or semi-device-independent regimes, where assumptions on the trustworthiness of measurements are relaxed. Another relevant line would be to explore how ping strategies can be integrated into entanglement routing algorithms, with real-time diagnostics feeding back into path selection or entanglement scheduling.

As a final remark, it is worthwhile to clarify that QPing strategies do not aim to substitute the classical ping program, which is a prerequisite for quantum networks as well, but to extend its utility to quantum connectivity. 
Indeed, quantum ping primitives provide a simple but powerful tool to dynamically certify entanglement-based connectivity in realistic quantum networks. We expect it to play a central role in the operational stack of future quantum communication architectures.

\section*{Acknowledgments} This research was funded in whole or in part by the Austrian Science Fund (FWF) 10.55776/P36009 and 10.55776/P36010. For open access purposes, the author has applied a CC BY public copyright license to any author-accepted manuscript version arising from this submission.
This work has been funded by the European Union under the ERC grant QNattyNet, n.101169850. Views and opinions expressed are however those of the author(s) only and do not necessarily reflect those of the European Union or the European Research Council. Neither the European Union nor the granting authority can be held responsible for them.




\bibliography{Qping}

\begin{thebibliography}{10}

\bibitem{Giovannetti_2011}
V.~Giovannetti, S.~Lloyd, and L.~Maccone, ``Advances in quantum metrology,''
  {\em Nature Photonics}, vol.~5, p.~222–229, Mar. 2011.

\bibitem{Kessler2014}
E.~M. Kessler, I.~Lovchinsky, A.~O. Sushkov, and M.~D. Lukin, ``Quantum error
  correction for metrology,'' {\em Phys. Rev. Lett.}, vol.~112, p.~150802, Apr
  2014.

\bibitem{Sekatski2020}
P.~Sekatski, S.~W\"olk, and W.~D\"ur, ``Optimal distributed sensing in noisy
  environments,'' {\em Phys. Rev. Research}, vol.~2, p.~023052, Apr 2020.

\bibitem{GiaWinCon-25}
A.~Giani, M.~Z. Win, and A.~Conti, ``Quantum sensing and communication via
  non-gaussian states,'' {\em IEEE Journal on Selected Areas in Information
  Theory}, vol.~6, pp.~18--33, 2025.

\bibitem{Gisin2002}
N.~Gisin, G.~Ribordy, W.~Tittel, and H.~Zbinden, ``Quantum cryptography,'' {\em
  Rev. Mod. Phys.}, vol.~74, pp.~145--195, Mar 2002.

\bibitem{Pirandola_2020}
S.~Pirandola, U.~L. Andersen, L.~Banchi, M.~Berta, D.~Bunandar, R.~Colbeck,
  D.~Englund, T.~Gehring, C.~Lupo, C.~Ottaviani, J.~L. Pereira, M.~Razavi,
  J.~Shamsul~Shaari, M.~Tomamichel, V.~C. Usenko, G.~Vallone, P.~Villoresi, and
  P.~Wallden, ``Advances in quantum cryptography,'' {\em Advances in Optics and
  Photonics}, vol.~12, p.~1012, Dec. 2020.

\bibitem{CiracDistributed}
J.~I. Cirac, A.~K. Ekert, S.~F. Huelga, and C.~Macchiavello, ``Distributed
  quantum computation over noisy channels,'' {\em Phys. Rev. A}, vol.~59,
  pp.~4249--4254, Jun 1999.

\bibitem{Hayashi15}
M.~Hayashi and T.~Morimae, ``Verifiable measurement-only blind quantum
  computing with stabilizer testing,'' {\em Phys. Rev. Lett.}, vol.~115,
  p.~220502, Nov 2015.

\bibitem{Cacciapuoti2020}
A.~S. Cacciapuoti, M.~Caleffi, F.~Tafuri, F.~S. Cataliotti, S.~Gherardini, and
  G.~Bianchi, ``Quantum internet: Networking challenges in distributed quantum
  computing,'' {\em {IEEE} Network}, vol.~34, pp.~137--143, jan 2020.

\bibitem{jensen2025quantum}
K.~S. Jensen, L.~Valentini, R.~B. Christensen, M.~Chiani, and P.~Popovski,
  ``Quantum two-way protocol beyond superdense coding: Joint transfer of data
  and entanglement,'' {\em IEEE Transactions on Quantum Engineering}, 2025.

\bibitem{Kimble2008}
H.~J. Kimble, ``The quantum internet,'' {\em Nature}, vol.~453, pp.~1023--1030,
  June 2008.

\bibitem{Wehner2018}
S.~Wehner, D.~Elkouss, and R.~Hanson, ``Quantum internet: A vision for the road
  ahead,'' {\em Science}, vol.~362, p.~eaam9288, oct 2018.

\bibitem{CacCalVan-20}
A.~S. Cacciapuoti, M.~Caleffi, R.~Van~Meter, and L.~Hanzo, ``When entanglement
  meets classical communications: Quantum teleportation for the quantum
  internet,'' {\em IEEE Transactions on Communications}, vol.~68, no.~6,
  pp.~3808--3833, 2020.

\bibitem{Gyongyosi_2022}
L.~Gyongyosi and S.~Imre, ``Advances in the quantum internet,'' {\em
  Communications of the ACM}, vol.~65, p.~52–63, July 2022.

\bibitem{Koji2023}
K.~Azuma, S.~E. Economou, D.~Elkouss, P.~Hilaire, L.~Jiang, H.-K. Lo, and
  I.~Tzitrin, ``Quantum repeaters: From quantum networks to the quantum
  internet,'' {\em Rev. Mod. Phys.}, vol.~95, p.~045006, Dec 2023.

\bibitem{rfc9583}
C.~Wang, A.~Rahman, R.~Li, M.~Aelmans, and K.~Chakraborty, ``{Application
  Scenarios for the Quantum Internet}.'' RFC 9583, June 2024.

\bibitem{Pirker_2018}
A.~Pirker, J.~Wallnöfer, and W.~Dür, ``Modular architectures for quantum
  networks,'' {\em New Journal of Physics}, vol.~20, p.~053054, May 2018.

\bibitem{Kozlowski2019}
W.~Kozlowski and S.~Wehner, ``Towards large-scale quantum networks,'' in {\em
  Proceedings of the Sixth Annual ACM International Conference on Nanoscale
  Computing and Communication}, NANOCOM '19, (New York, NY, USA), Association
  for Computing Machinery, 2019.

\bibitem{Kozlowski_2020}
W.~Kozlowski, A.~Dahlberg, and S.~Wehner, ``Designing a quantum network
  protocol,'' in {\em Proceedings of the 16th International Conference on
  emerging Networking EXperiments and Technologies}, CoNEXT ’20, p.~1–16,
  ACM, Nov. 2020.

\bibitem{Pirker_2019}
A.~Pirker and W.~Dür, ``A quantum network stack and protocols for reliable
  entanglement-based networks,'' {\em New Journal of Physics}, vol.~21,
  p.~033003, Mar. 2019.

\bibitem{Dahlberg_2019}
A.~Dahlberg, M.~Skrzypczyk, T.~Coopmans, L.~Wubben, F.~Rozpędek, M.~Pompili,
  A.~Stolk, P.~Pawełczak, R.~Knegjens, J.~de~Oliveira~Filho, R.~Hanson, and
  S.~Wehner, ``A link layer protocol for quantum networks,'' in {\em
  Proceedings of the ACM Special Interest Group on Data Communication}, SIGCOMM
  ’19, p.~159–173, ACM, Aug. 2019.

\bibitem{Illiano_2022}
J.~Illiano, M.~Caleffi, A.~Manzalini, and A.~S. Cacciapuoti, ``Quantum internet
  protocol stack: A comprehensive survey,'' {\em Computer Networks}, vol.~213,
  p.~109092, Aug. 2022.

\bibitem{Briegel_Repeaters}
H.-J. Briegel, W.~D\"ur, J.~I. Cirac, and P.~Zoller, ``Quantum repeaters: The
  role of imperfect local operations in quantum communication,'' {\em Phys.
  Rev. Lett.}, vol.~81, pp.~5932--5935, Dec 1998.

\bibitem{D_r_2016}
W.~Dür, H.~Briegel, P.~Zoller, and P.~van Loock, ``Quantum repeater,'' in {\em
  Quantum Information: From Foundations to Quantum Technology Applications}
  (D.~Bruß and G.~Leuchs, eds.), p.~691–700, Wiley, April 2016.

\bibitem{Meter_2013}
R.~Meter and J.~Touch, ``Designing quantum repeater networks,'' {\em IEEE
  Commun. Mag.}, vol.~51, p.~64–71, August 2013.

\bibitem{MR2023}
J.~Miguel-Ramiro, F.~Riera-S\`abat, and W.~D\"ur, ``Quantum repeater for $w$
  states,'' {\em PRX Quantum}, vol.~4, p.~040323, Nov 2023.

\bibitem{Van_Meter_2013}
R.~Van~Meter, T.~Satoh, T.~D. Ladd, W.~J. Munro, and K.~Nemoto, ``Path
  selection for quantum repeater networks,'' {\em Networking Science}, vol.~3,
  p.~82–95, Dec. 2013.

\bibitem{Lee_2022}
Y.~Lee, E.~Bersin, A.~Dahlberg, S.~Wehner, and D.~Englund, ``A quantum router
  architecture for high-fidelity entanglement flows in quantum networks,'' {\em
  npj Quantum Information}, vol.~8, p.~75, June 2022.

\bibitem{Shi_2024}
W.~Shi and R.~Malaney, ``Quantum routing for emerging quantum networks,'' {\em
  IEEE Network}, vol.~38, p.~140–146, Jan. 2024.

\bibitem{Abane_2025}
A.~Abane, M.~Cubeddu, V.~S. Mai, and A.~Battou, ``Entanglement routing in
  quantum networks: A comprehensive survey,'' {\em IEEE Transactions on Quantum
  Engineering}, vol.~6, p.~1–39, 2025.

\bibitem{Meignant2019}
C.~Meignant, D.~Markham, and F.~Grosshans, ``Distributing graph states over
  arbitrary quantum networks,'' {\em Phys. Rev. A}, vol.~100, p.~052333, Nov
  2019.

\bibitem{Navascues2020}
M.~Navascu\'es, E.~Wolfe, D.~Rosset, and A.~Pozas-Kerstjens, ``Genuine network
  multipartite entanglement,'' {\em Phys. Rev. Lett.}, vol.~125, p.~240505, Dec
  2020.

\bibitem{LiXueLi-23}
Z.~Li, K.~Xue, J.~Li, L.~Chen, R.~Li, Z.~Wang, N.~Yu, D.~S.~L. Wei, Q.~Sun, and
  J.~Lu, ``Entanglement-assisted quantum networks: Mechanics, enabling
  technologies, challenges, and research directions,'' {\em IEEE Communications
  Surveys \& Tutorials}, vol.~25, no.~4, pp.~2133--2189, 2023.

\bibitem{Miguel_Ramiro_2023}
J.~Miguel-Ramiro, A.~Pirker, and W.~Dür, ``Optimized quantum networks,'' {\em
  Quantum}, vol.~7, p.~919, Feb. 2023.

\bibitem{Fan2024}
X.~Fan, C.~Zhan, H.~Gupta, and C.~R. Ramakrishnan, ``Optimized distribution of
  entanglement graph states in quantum networks,'' {\em arXiv:2405.00222},
  2024.

\bibitem{KurRos-12}
J.~F. Kurose and K.~W. Ross, {\em Computer Networking: A Top-Down Approach (6th
  Edition)}.
\newblock Pearson, 6th~ed., 2012.

\bibitem{rfc1122}
R.~T. Braden, ``{Requirements for Internet Hosts - Communication Layers}.'' RFC
  1122, Oct. 1989.

\bibitem{Wei_2022}
S.~Wei, B.~Jing, X.~Zhang, J.~Liao, C.~Yuan, B.~Fan, C.~Lyu, D.~Zhou, Y.~Wang,
  G.~Deng, H.~Song, D.~Oblak, G.~Guo, and Q.~Zhou, ``Towards real‐world
  quantum networks: A review,'' {\em Laser amp; Photonics Reviews}, vol.~16,
  Jan. 2022.

\bibitem{Azuma23}
K.~Azuma, S.~E. Economou, D.~Elkouss, P.~Hilaire, L.~Jiang, H.-K. Lo, and
  I.~Tzitrin, ``Quantum repeaters: From quantum networks to the quantum
  internet,'' {\em Rev. Mod. Phys.}, vol.~95, p.~045006, Dec 2023.

\bibitem{Vargas2021}
E.~Mart\'{\i}nez~Vargas, C.~Hirche, G.~Sent\'{\i}s, M.~Skotiniotis, M.~Carrizo,
  R.~Mu\~noz Tapia, and J.~Calsamiglia, ``Quantum sequential hypothesis
  testing,'' {\em Phys. Rev. Lett.}, vol.~126, p.~180502, May 2021.

\bibitem{Pallister2018}
S.~Pallister, N.~Linden, and A.~Montanaro, ``Optimal verification of entangled
  states with local measurements,'' {\em Phys. Rev. Lett.}, vol.~120,
  p.~170502, Apr 2018.

\bibitem{Yu_2019}
X.-D. Yu, J.~Shang, and O.~Gühne, ``Optimal verification of general bipartite
  pure states,'' {\em npj Quantum Information}, vol.~5, p.~112, Dec. 2019.

\bibitem{MiguelRamiro2022}
J.~Miguel-Ramiro, F.~Riera-S\`abat, and W.~D\"ur, ``Collective operations can
  exponentially enhance quantum state verification,'' {\em Phys. Rev. Lett.},
  vol.~129, p.~190504, Oct 2022.

\bibitem{Fields2024}
G.~Fields, N.~Sangwan, J.~Postlewaite, S.~Guha, and T.~Javidi, ``Sequential
  hypothesis testing of quantum states,'' in {\em 2024 IEEE Information Theory
  Workshop (ITW)}, p.~372–377, IEEE, Nov. 2024.

\bibitem{MiguelRamiro2023}
F.~Riera-S\`abat, J.~Miguel-Ramiro, and W.~D\"ur, ``Nondestructive verification
  of entangled states via fidelity witnessing,'' {\em Phys. Rev. A}, vol.~107,
  p.~022414, Feb 2023.

\bibitem{G_hne_2009}
O.~Gühne and G.~Tóth, ``Entanglement detection,'' {\em Physics Reports},
  vol.~474, p.~1–75, Apr. 2009.

\bibitem{Mohseni2008}
M.~Mohseni, A.~T. Rezakhani, and D.~A. Lidar, ``Quantum-process tomography:
  Resource analysis of different strategies,'' {\em Phys. Rev. A}, vol.~77,
  p.~032322, Mar 2008.

\bibitem{Cramer_2010}
M.~Cramer, M.~B. Plenio, S.~T. Flammia, R.~Somma, D.~Gross, S.~D. Bartlett,
  O.~Landon-Cardinal, D.~Poulin, and Y.-K. Liu, ``Efficient quantum state
  tomography,'' {\em Nature Communications}, vol.~1, p.~149, Dec. 2010.

\bibitem{rfc792}
``{Internet Control Message Protocol}.'' RFC 792, Sept. 1981.

\bibitem{pompili2022}
M.~Pompili, C.~Delle~Donne, I.~te~Raa, B.~van~der Vecht, M.~Skrzypczyk,
  G.~Ferreira, L.~de~Kluijver, A.~J. Stolk, S.~L. Hermans, P.~Pawe{\l}czak,
  {\em et~al.}, ``Experimental demonstration of entanglement delivery using a
  quantum network stack,'' {\em npj Quantum Information}, vol.~8, no.~1,
  p.~121, 2022.

\bibitem{nielsen_chuang_2010}
M.~A. Nielsen and I.~L. Chuang, {\em \emph{Quantum Computation and Quantum
  Information: 10th Anniversary Edition}}.
\newblock Cambridge University Press, 2010.

\bibitem{Meter2013b}
R.~{Van Meter}, T.~Satoh, T.~D. Ladd, W.~J. Munro, and K.~Nemoto, ``{Path
  selection for quantum repeater networks},'' {\em Networking Science}, vol.~3,
  pp.~82--95, Dec 2013.

\bibitem{Gyongyo2017}
L.~Gyongyosi and S.~Imre, ``Entanglement-gradient routing for quantum
  networks,'' {\em Scientific Reports}, vol.~7, no.~1, p.~14255, 2017.

\bibitem{Gyong2018}
L.~Gyongyosi and S.~Imre, ``Decentralized base-graph routing for the quantum
  internet,'' {\em Phys. Rev. A}, vol.~98, p.~022310, Aug 2018.

\bibitem{Pant2019}
M.~Pant, H.~Krovi, D.~Towsley, L.~Tassiulas, L.~Jiang, P.~Basu, D.~Englund, and
  S.~Guha, ``Routing entanglement in the quantum internet,'' {\em npj Quantum
  Information}, vol.~5, no.~1, p.~25, 2019.

\bibitem{delocalizedinfo}
J.~Miguel-Ramiro and W.~Dür, ``Delocalized information in quantum networks,''
  {\em New J. Phys.}, vol.~22, p.~043011, apr 2020.

\bibitem{Yu2021}
N.~Yu, C.-Y. Lai, and L.~Zhou, ``Protocols for packet quantum network
  intercommunication,'' {\em IEEE Transactions on Quantum Engineering}, vol.~2,
  p.~1–9, 2021.

\bibitem{DaiRinTow-21}
W.~Dai, A.~Rinaldi, and D.~Towsley, ``Entanglement swapping in quantum
  switches: Protocol design and stability analysis,'' {\em arXiv preprint
  arXiv:2110.04116}, 2021.

\bibitem{Freund2024}
J.~Freund, A.~Pirker, and W.~D\"ur, ``Flexible quantum data bus for quantum
  networks,'' {\em Phys. Rev. Res.}, vol.~6, p.~033267, Sep 2024.

\bibitem{Tan-10}
A.~S. Tanenbaum and D.~J. Wetherall, {\em Computer Networks}.
\newblock Prentice Hall Press, 5th~ed., 2010.

\bibitem{StandForms}
W.~D\"ur, M.~Hein, J.~I. Cirac, and H.-J. Briegel, ``Standard forms of noisy
  quantum operations via depolarization,'' {\em Phys. Rev. A}, vol.~72,
  p.~052326, Nov 2005.

\bibitem{RandoCompiling}
A.~Hashim, R.~K. Naik, A.~Morvan, J.-L. Ville, B.~Mitchell, J.~M. Kreikebaum,
  M.~Davis, E.~Smith, C.~Iancu, K.~P. O'Brien, I.~Hincks, J.~J. Wallman,
  J.~Emerson, and I.~Siddiqi, ``Randomized compiling for scalable quantum
  computing on a noisy superconducting quantum processor,'' {\em Phys. Rev. X},
  vol.~11, p.~041039, Nov 2021.

\bibitem{Dahlberg2018}
A.~Dahlberg and S.~Wehner, ``Transforming graph states using single-qubit
  operations,'' {\em Philosophical Transactions of the Royal Society A:
  Mathematical, Physical and Engineering Sciences}, vol.~376, p.~20170325, May
  2018.

\bibitem{Dahlberg2020}
A.~Dahlberg, J.~Helsen, and S.~Wehner, ``Transforming graph states to
  bell-pairs is {NP}-complete,'' {\em Quantum}, vol.~4, p.~348, Oct. 2020.

\bibitem{Hein2004}
M.~Hein, J.~Eisert, and H.~J. Briegel, ``Multiparty entanglement in graph
  states,'' {\em Phys. Rev. A}, vol.~69, p.~062311, Jun 2004.

\bibitem{Hein2006}
M.~Hein, W.~Dür, J.~Eisert, R.~Raussendorf, M.~Nest, and H.~Briegel,
  ``Entanglement in graph states and its applications,'' {\em in Quantum
  Computers, Algorithms and Chaos, Proceedings of the International School of
  Physics “Enrico Fermi,” Vol. 162, Varenna, 2005, edited by G. Casati, D.
  L. Shepelyansky, P. Zoller, and G. Benenti (IOS Press, Amsterdam}, vol.~162,
  pp.~115--218, 03 2006.

\bibitem{MorRuiz2024}
M.~F. Mor-Ruiz and W.~D\"{u}r, ``Influence of noise in entanglement-based
  quantum networks,'' {\em IEEE Journal on Selected Areas in Communications},
  vol.~42, p.~1793–1807, July 2024.

\bibitem{MorRuiz2025}
M.~F. Mor-Ruiz, J.~Walln\"{o}fer, and W.~D\"{u}r, ``Imperfect quantum networks
  with tailored resource states,'' {\em Quantum}, vol.~9, p.~1605, Jan. 2025.

\bibitem{Bartkiewicz2013}
K.~Bartkiewicz, B.~Horst, K.~Lemr, and A.~Miranowicz, ``Entanglement estimation
  from bell inequality violation,'' {\em Phys. Rev. A}, vol.~88, p.~052105, Nov
  2013.

\bibitem{Friis2018}
N.~Friis, G.~Vitagliano, M.~Malik, and M.~Huber, ``Entanglement certification
  from theory to experiment,'' {\em Nature Reviews Physics}, vol.~1,
  p.~72–87, Dec. 2018.

\bibitem{Brandao2005}
F.~G. S.~L. Brand\~ao, ``Quantifying entanglement with witness operators,''
  {\em Phys. Rev. A}, vol.~72, p.~022310, Aug 2005.

\bibitem{Hyllus2005}
P.~Hyllus, O.~G\"uhne, D.~Bru\ss{}, and M.~Lewenstein, ``Relations between
  entanglement witnesses and bell inequalities,'' {\em Phys. Rev. A}, vol.~72,
  p.~012321, Jul 2005.

\bibitem{Eisert2020}
J.~Eisert, D.~Hangleiter, N.~Walk, I.~Roth, D.~Markham, R.~Parekh, U.~Chabaud,
  and E.~Kashefi, ``Quantum certification and benchmarking,'' {\em Nature
  Reviews Physics}, vol.~2, p.~382–390, June 2020.

\end{thebibliography}

\bibliographystyle{ieeetr}

\clearpage


\end{document}